\documentstyle[epsfig,12pt]{article}
\let\section=\subsection     \let\subsection=\subsubsection

\newcommand{\be}{\begin{equation}}
\newcommand{\ee}{\end{equation}}
\newcommand{\bd}{\begin{displaymath}}
\newcommand{\ed}{\end{displaymath}}
\newcommand{\ba}{\begin{eqnarray}}
\newcommand{\ea}{\end{eqnarray}}

\newcommand{\ave}[1]{\langle {#1} \rangle}

\begin{document}
\begin{center}
   {\large \bf DROPPING $\sigma$-MESON MASS}\\[2mm]
   {\large \bf AND IN-MEDIUM S-WAVE $\pi\pi$ CORELLATIONS}\\[5mm]
    P. SCHUCK$^1$, Z. AOUISSAT$^2$, G. CHANFRAY$^3$, J. WAMBACH$^2$\\[5mm]
   {\small \it $^1$  ISN,Universit\'e Joseph Fourier, CNRS-IN2P3, \\ 53,av des Martyrs,
F-38026 Grenoble C\'edex,France. \\$^2$  IKP, Technische Universit\"at 
Darmstadt, \\ Schlo{\ss}gartenstra{\ss}e 9, 64289 Darmstadt, Germany.\\
$^3$   IPN Lyon. \\ 43 Bd. du 11 Novembre 1918,  F69622 Villeurbanne
C\'edex, France.   \\[8mm] }
\end{center}

\begin{abstract}\noindent
The influence of a dropping $\sigma$-meson mass on previously 
calculated in-medium $\pi\pi$ correlations in the 
$J=I=0$ ($\sigma$-meson) channel \cite{borm, arcsw} is investigated. 
It is found that the invariant-mass distribution around the vacuum threshold
experiences a further strong enhancement over and above the standard many-body 
effects. The relevance of this result for the theoretical interpretation of 
recent $A(\pi,2\pi)X$ data is pointed out.

\end{abstract}
\vspace{0.6cm}





\section{Introduction}
It is about ten years now that the first theoretical work suggesting a strong 
enhancement around the $2m_{\pi}$ threshold of the in-medium 'sigma'-meson mass 
distribution was published \cite{snc}. The threshold enhancement originates from 
in-matter p-wave renormalization of the two pions to which the 'bare sigma' meson 
of about 1 GeV decays. The existence of such a bare sigma meson has in the meanwhile 
been confirmed by quenched lattice calculations \cite{pch}. Its decay into two pions 
leads to a strong hybridization of the sigma meson and finally only a broad bump of 
about 600 MeV width survives at considerably lower energy. The $\pi\pi$ interaction
constants in the completely phenomenological model of ref.~\cite{snc} had been adjusted
to reproduce the experimental $\pi\pi$ phase shift in the scalar -isoscalar channel.
Since the latter raises right from threshold with considerable positive slope, it 
signals substantial attraction among low-energy pions. Pions are, however, too light 
to be bound by the interaction. On the other hand, the p-wave renormalization of the 
pions in the nuclear medium (nucleon- and delta-hole couplings) increases their kinetic 
mass and thus suppresses the kinetic energy. As a consequence binding of the 
two pions is favored at high density. Before this occurs, a strong accumulation of 
strength close to the $2m_{\pi}$ threshold takes place. Even if, with increasing density,
the mass distribution receives contributions from below the $2m_{\pi}$ threshold, this 
does not lead to a sharp 2-pion bound state. Various decay channels such as $\pi N$-
$\pi\Delta$- or $NN$ render it broad \cite{arcsw}. To describe the subthreshold 
behavior of the scalar-isoscalar strength function in accordance with 
chiral symmetry requirements it turned out later that the $\pi\pi$ scattering needs
to be implemented by using chiral models such as the linear or non-linear sigma model,
rather than an empirical parametrization of the s-wave phase shift.
In addition a formalism is called for which respects chiral symmetry beyond the tree level,  
\cite{asw}. Otherwise the in-medium 
effects become so strong that they lead to unreasonable pion-pair condensation \cite{modph}.
In the meantime s-wave pion-pion correlations have attracted a significant amount of 
attention both on the theoretical \cite{cov,vicent,hat} and the experimental \cite{bonu,ng} 
sides. This is primarily related to the fact that these studies are of relevance for 
the evolution of the chiral condensate and its fluctuations with increasing density.
The enhancement of the in-medium $\pi\pi$ correlations close to and below the 
vacuum threshold has recently been confirmed by an independent calculation within the 
non-linear sigma model. Also on the experimental side important progress has 
been made. The CHAOS collaboration has studied the $A(\pi,2\pi)$ knock-out 
reaction \cite{bonu,ng} and the invariant-mass measurements of the outgoing pions 
reveal a strong low-mass enhancement which seems to corroborate the theoretical predictions.
Performing elaborate reaction calculations, including initial- and final-state absorption, 
Vincente-Vacas and Oset\cite{vicent} have claimed that the theory outlined above
underestimates the measured $\pi\pi$ mass enhancement. This claim may be
partly questioned, since the reaction theory calls for the inclusion with a finite 
total three momentum of the in-medium pion pairs which sofar are only
propagated in back-to-back kinematics ($\vec q=0$). It was shown in \cite{borm} that,
allowing for finite three momenta of the pion pair, leads to further increase the 
$\pi^+\pi^-$ invariant-mass distribution near threshold. 
On the other hand, Hatsuda et al. \cite{hat} have argued that the partial restoration 
of chiral symmetry in nuclear matter, which leads to a dropping of the $\sigma$-meson 
mass \cite{br}, induces similar effects as the standard many-body correlation mentioned 
above. It is therefore natural to study the combination of both effects.
 
\section{The Model}

As a model for $\pi\pi$ scattering we consider the linear sigma model treated in 
leading order of the $1/N$-expansion \cite{asw}.
The scattering matrix can then be cast in the following form
\begin{eqnarray}
T_{ab,cd}(s) \,&=& \delta_{ab}\delta_{cd} 
\frac{D_{\pi}^{-1}(s) - D_{\sigma}^{-1}(s)}{3\ave{\sigma}^2}
\, \frac{D_{\sigma}(s)}{D_{\pi}(s)}~,
\label{eq1}
\end{eqnarray}
where $s$ is the Mandelstam variable. In Eq.~(\ref{eq1}) $D_{\pi}(s)$ and $D_{\sigma}(s)$ 
denote respectively the full pion and sigma propagators,
while $\ave{\sigma}$ is the sigma condensate. The expression in Eq.~(\ref{eq1}) 
reduces, in the soft-pion limit, to a Ward identity which links the $\pi\pi$ 
four-point function to the $\pi$ and $\sigma$ two-point functions as well
as to the $\sigma$ one-point function.
To this order, the pion propagator and the sigma-condensate are obtained
from a Hartree-Bogoliubov approximation \cite{asw}. 
In terms of the pion mass $m_{\pi}$ and the decay constant 
$f_{\pi}$, they are given as  
\begin{equation}
D_{\pi}(s) = \frac{1}{s - m_{\pi}^2}, \quad\quad f_{\pi} \,=\,\sqrt{3} \ave{\sigma}.
\label{eq2}
\end{equation} 
The sigma meson, on the other hand, is obtained from the Random-Phase Approximation 
(RPA) involving $\pi\pi$ scattering \cite{asw} and reads    
\begin{equation}
D_{\sigma}(s) \,=\, \left[{ s\,-\, m_{\sigma}^2
\,-\, \frac{2 \lambda^4 \ave{\sigma}^2\,{\Sigma}_{\pi\pi}(s)}
{ 1\,-\,  \lambda^2 {\Sigma}_{\pi\pi}(s)}}\right]^{-1}~,
 \label{eq3}
\end{equation}
where ${\Sigma}_{\pi\pi}(s)$ is the $\pi\pi$ self energy, regularized by means 
of a form factor which is used as a fit function \cite{borm} and allows 
to reproduce the experimental $\pi\pi$ phase shifts. The coupling constant 
$\lambda^2$ denotes the bare quartic coupling of the linear $\sigma$-model, 
related to the mean-field pion mass, $m_{\pi}$, the sigma mass, $m_{\sigma}$, 
and the condensate, $\ave{\sigma}$, via the mean-field saturated  Ward identity
 
\begin{equation}
m_{\sigma}^2 = m_{\pi}^2 + 2 \lambda^2\ave{\sigma}^2. 
 \label{eq4}
\end{equation}
It is clear from the above that the sigma-meson propagator in this 
approach is correctly defined, since it satisfies a hierarchy of Ward 
identities.

In cold nuclear matter the pion is dominantly coupled to $\Delta -h$,  $p-h$, as 
well as to $2p-2h$ excitations which, on the other hand, 
are renormalized by means of repulsive nuclear short-range correlations,   
(see \cite{arcsw} for details). 
Since the pion is a (near) Goldstone mode,  its in-medium s-wave renormalization 
does not induce appreciable changes. The sigma meson, on the other hand,
is not protected by chiral symmetry from large s-wave renormalization, resulting
in a density-dependent mass modification \cite{pch} . We extract an approximate 
density dependence 
at mean-field level from the (leading-order) density dependence of
the condensate. Indeed from Eq.~(\ref{eq4}) it is clear that the density dependence
of the sigma meson mass is essentially dictated by the density dependence of the 
condensate.

\begin{figure}[htb]
\centerline{ 
\epsfig{file=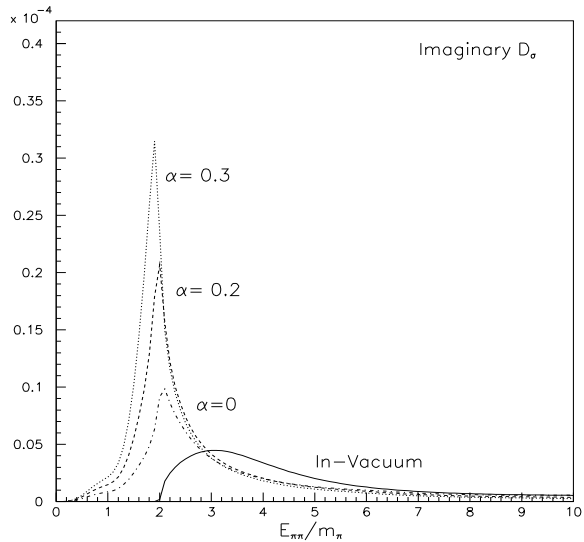,width=8.6cm,height=8.cm,angle=0}}
\caption[fig1]{\small
Results for the imaginary part of the in-medium sigma-meson propagator. 
Except for the vacuum case (full line curve) the remaining in-medium curves are 
computed at normal nuclear matter density. The dashed-dotted curve is for $\alpha=0$, 
dashed for $\alpha=0.2$ and the dotted for $\alpha=0.3$.
\label{fig1.} }
\end{figure}

Thus, for densities below and around nuclear saturation density, $\rho_0$, we take 
for the in-medium sigma-meson mass the simple ansatz
\begin{equation} 
m_{\sigma}(\rho)= m_{\sigma}(1 -\alpha \frac{\rho}{\rho_0})
\label{eq5}
\end{equation} 
where $\rho$ is the nuclear matter density and $m_{\sigma}$ is the vacuum  
$\sigma$-meson mass. Such a density dependence very naturally arises in the linear 
sigma model from the tad-pole graph where the sigma meson directly couples to the nuclear 
density. In \cite{hat} this density dependence was investigated quantitatively and a value 
of $\alpha$ in the range of 0.2 to 0.3 was found. These are the values which we also will use.
It should be mentioned at this point that between the (s-wave) density renormalization via 
the tad-pole of the bare sigma-meson mass in the linear sigma model (as we mentioned already,
such a bare sigma exists in quenched QCD lattice calculations \cite{pch}) and the in-medium 
p-wave renormalization of the pion there exists no double counting. Besides 
the different partial waves involved, a simple inspection of the corresponding Feynman graphs 
shows that both density effects are of totally different microscopic origin.
     
\section{Results}

The result for the invariant-mass distribution $Im D_{\sigma}(E_{\pi\pi})$, as
calculated from Eq.~(3) by using the in-medium mass (\ref{eq5}), is shown in 
Fig.~1 at saturation density. One observes a dramatic downward shift of the mass 
distribution as compared to the vacuum. The low-energy enhancement,
already present without sigma-mass modification ($\alpha=0$) and induced by the 
density-dependence of the pion loop,
is strongly reinforced as the in-medium $\sigma$-meson mass is included. 
For $\alpha=0.2$ and  $\alpha=0.3$ the peak height is increased by a factor 2 and 4 
respectively. Similarly for the T-matrix, a sizable effect can be noticed in its 
imaginary part, as shown in Fig.~2, which might be sufficient 
to explain the findings of the CHAOS collaboration \cite{bonu}. Further work in this 
direction is in progress.

\begin{figure}[htb]
\centerline{ 
\epsfig{file=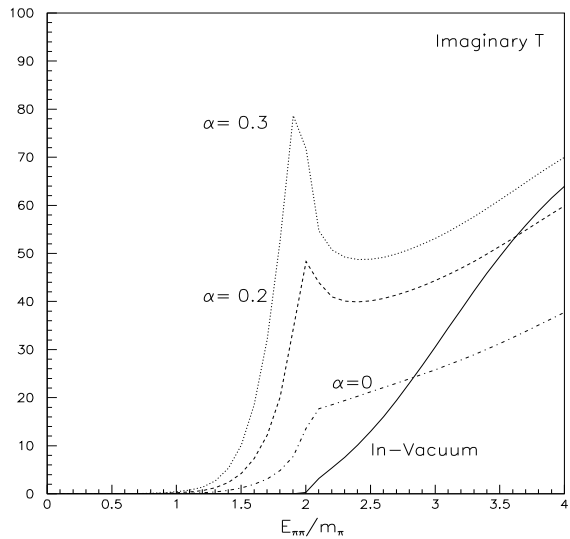,width=8.6cm,height=8.cm,angle=0}}
\caption[fig1]{\small
Results for the imaginary part of the in-medium 
T-matrix for $\pi\pi$ scattering. 
Except for the vacuum case (full line curve) 
the remaining in-medium curves are 
computed at normal nuclear matter density. 
 The dashed-dotted curve is for $\alpha=0$, the dashed 
for $\alpha=0.2$ and the dotted for $\alpha=0.3$.
\label{fig2.} }
\end{figure}

\section{Discussions and Conclusion}

Let us further comment and discuss the above results. They clearly indicate that there exists 
a large conspiracy between s- and p-wave in-medium renormalisations of the hybrid 'sigma meson'.
Both effects induce a strong shift of the strength towards and even below the $2m_{\pi}$ 
threshold. However, one has to worry about opposing effects. For example it may happen 
that vertex corrections, usually a source of repulsion and not taken into account in this 
work, weaken the effects. They seem, however, to be of minor importance as was recently 
shown by Chanfray et al.~\cite{dd}. More care should also be taken in properly incorporating 
Pauli-blocking  when renormalizing the pion pairs in matter, although preliminary 
investigations \cite{thesis} have shown that this effect is weak.

In conclusion we have shown that a dropping sigma-meson mass, linked to the partial restoration 
of chiral symmetry in nuclear matter, further enhances the build up of previously found 
$\pi\pi$ strength in the $I = J = 0$ channel. This scenario holds in the linear sigma model 
but is likely to hold true in the non-linear sigma model as well, since the tad-pole graph 
which renormalizes the bare sigma meson mass in the linear sigma model has its equivalent in 
the non-linear version. Whether our findings are linked to similar ones found recently 
in the experiment by Bonutti et al \cite{bonu} is an exciting possibility which, however, must 
still be consolidated by refined reaction calculations. Last but not least there remains the 
challenge that the results of Bonutti et al. are confirmed by other experiments. A promising 
reaction will be the $A(\gamma,2\pi)$ reaction off nuclei with varying mass number, 
since the average density increases from low to high mass numbers and photons,
in contrast to pions, probe the nuclear interior. Existing data from MAMI Mainz \cite{main} 
are waiting to be analyzed! We hope to have shown in this contribution \cite{epja} that the 
scalar-isoscalar channel, in what concerns chiral symmetry, is at least as interesting and 
promising as the vector-isovector channel (rho-meson).
\vspace{0.5cm}

\noindent{\bf Acknowledgements:}

\noindent
Fruitful and stimulating discussions with N. Grion, W. N\"orenberg, E. Oset, 
M. Vincente-Vacas, T. Walcher, and W. Weise are appreciated.

\end{document}